\documentclass[12pt,preprint]{aastex}

\usepackage{apjfonts}
\usepackage{url}
\usepackage{hyperref}
\usepackage{amsmath}

\usepackage{color}
\definecolor{dark-red}{rgb}{0.8,0,0}
\definecolor{dark-green}{rgb}{0,0.4,0}
\definecolor{dark-blue}{rgb}{0,0,0.8}
\definecolor{dark-margenta}{rgb}{0.8,0,0.8}
\definecolor{orange}{rgb}{1.0,0.6,0}
\newcommand{\hR}{\color{black}}
\newcommand{\hG}{\color{black}}
\newcommand{\hB}{\color{black}}
\newcommand{\hM}{\color{black}}
\newcommand{\hO}{\color{black}}
\DeclareMathOperator{\slog}{slog}
\DeclareMathOperator{\sign}{sign}
\newcommand{\PS}{$ \mathrel{\rotatebox[origin=c]{-90}{$\subset$} \!\!\!\! \raisebox{.2em}{$\mid$}}\: $}
\newcommand{\N}[1]{${\rm N}_{#1}$}
\newcommand{\M}[1]{${\rm M}_{#1}$}
\newcommand{\CH}[1]{${\rm CH}_{#1}$}
\newcommand{\SC}[1]{${\rm SC}_{#1}$}
\newcommand{\BP}[1]{${\rm BP}_{#1}$}
\received{}
\revised{}
\accepted{}
\ccc{}
\cpright{}{}

\slugcomment{To appear in ApJ in November 2012}

\shorttitle{2010 August 1--2 Sympathetic Eruptions: I.}
\shortauthors{TITOV et al.}

\begin{document}

\title{2010 AUGUST 1--2 SYMPATHETIC ERUPTIONS: \\ 
I. MAGNETIC TOPOLOGY OF THE SOURCE-SURFACE BACKGROUND FIELD} 


\author{V. S. Titov\altaffilmark{1}, Z. Mikic\altaffilmark{1}, T. T\"{o}r\"{o}k\altaffilmark{1}, J. A. Linker\altaffilmark{1}}

\and

\author{O. Panasenco\altaffilmark{2}}

\altaffiltext{1}{Predictive Science Inc., 9990 Mesa Rim Road, Suite 170, San Diego, CA 92121} 

\altaffiltext{2}{Helio Research, 5212 Maryland Avenue, La Crescenta, CA 91214} 

\email{titovv@predsci.com}




\begin{abstract}

A sequence of apparently coupled eruptions was observed on 2010 August 1--2  by SDO and STEREO. 
The eruptions were closely synchronized with one another, even though some of them occurred at widely separated locations.
In an attempt to identify a plausible reason for such synchronization, we study the large-scale structure of the background magnetic configuration.
The coronal field was computed from the photospheric magnetic field observed at the appropriate time period by using the potential field source-surface model.

We investigate the resulting field structure by analyzing the so-called squashing factor calculated at the photospheric and source-surface boundaries, as well as at different coronal cross-sections.
Using this information as a guide, we determine the underlying structural skeleton of the configuration, including separatrix and quasi-separatrix surfaces. 
Our analysis reveals, in particular, several pseudo-streamers in the regions where the eruptions occurred.
Of special interest to us are the magnetic null points and separators associated with the pseudo-streamers.
We propose that magnetic reconnection triggered along these separators by the first eruption likely played a key role in establishing the assumed link between the sequential eruptions.
The present work substantiates our recent simplified magnetohydrodynamic model of sympathetic eruptions and provides a guide for further deeper study of these phenomena.
Several important implications of our results for the S-web model of the slow solar wind are also addressed.
\end{abstract}


\section{INTRODUCTION}
 \label{introduction}

{\hB
Coronal mass ejections (CMEs) are spectacular solar phenomena that have been intensely studied over more than forty years.
Being the main driver of space weather disturbances near the Earth, they are part of a more general eruption process, often including filament eruptions and flares.
Although it is now understood that these phenomena are due to a local destabilization of the coronal magnetic field,  many basic questions on the physics of CMEs are still under study \citep[e.g.,][]{Forbes2000b, Forbes2010}. 
Accordingly, theoretical and numerical investigations of CME initiation and 
evolution have so far focused mainly on {\it single} eruptions.

However, there also exist  {\it multiple} eruptions occurring within a 
relatively short period of time and at different, often widely separated, locations. In the largest
events, the respective source regions can cover a full hemisphere \citep[so-called ``global 
CMEs''; e.g.,][]{Zhukov2007}, so that such events naturally produce large heliospheric disturbances.
While it has been argued whether or not the temporal correlation of multiple eruptions  
is coincidental \citep[e.g.,][]{Biesecker2000}, both statistical investigations 
\citep[e.g.,][]{Moon2002, Wheatland2006} and detailed case studies \citep[e.g.,][]{Wang2001, Jiang2011, Yang2012, Shen2012} 
indicate that there are {\it causal connections} between them\footnote{We do not distinguish 
here between sympathetic flares and sympathetic CMEs, since typically both are part of the same eruption 
process.}. We accept this fact as a starting point of our study and will henceforth call such 
eruptions {\it sympathetic} or {\it linked}.

The physical mechanisms of these connections, however,  have yet to be unveiled.
In earlier works they have been related, for instance, to
destabilization by chromospheric large-scale waves \citep[][]{Ramsey1966} or 
large-scale properties of convective flows \citep[][]{Bumba1993}.
More recent research suggests that the mechanisms linking sympathetic eruptions act 
in the corona and involve its magnetic field structure. 
For example, one proposed mechanism relies on perturbations propagating along field lines between the source regions of eruptions \citep[e.g.,][]{Jiang2008}, 
while another appeals to changes in the background field due to reconnection 
\citep[][]{Liu2009a, Zuccarello2009b, Jiang2011, Shen2012}.
Yet such explanations were often based on qualitative and sometimes rather speculative considerations. 
}

The high-cadence, full-disk observations by SDO, along with studies of the large-scale coronal 
magnetic field, now provide us the opportunity to substantially increase our understanding of 
such eruptions. 
The event under study attracted considerable attention in the solar community and 
beyond.
%
It involved an entire hemisphere of the Sun, consisted of 
several flares and six filament eruptions and CMEs, and triggered 
a geomagnetic storm on August 3 \citep{Harrison2012}.  A detailed account of all eruptions 
and their precursors can be found in \cite{Schrijver2011}. Here we restrict ourselves to the 
main five eruptions, whose connections we aim to explain in the present study. 
Using a combination of SDO data and analysis of field line connectivity for the 2010 August 1--2 eruptions, \cite{Schrijver2011} found evidence that {\it all} involved source regions were connected by structural features such as separatrix surfaces, separators, and quasi-separatrix layers \citep[QSLs,][]{Priest1995, Demoulin1996, Titov2002}.  We have recently performed a simplified magnetohydrodynamic (MHD) simulation of a subset of these eruptions \citep{Torok2011}, in which two successive eruptions were initiated by reconnection at a separator high in the corona. 
Thus this work strongly supports the idea that the structural features can indeed play a key role in generating linked eruptions.    

While these new results are very encouraging, further research is needed. 
First, our simulation used only a simplified magnetic configuration and addressed just a subset of the complex sequence of CMEs on 2010 August 1--2.
Second, the findings by \cite{Schrijver2011}, being of a general nature, did not explain the exact role of structural features in connecting individual eruptions. 
We show here that a comprehensive structural analysis of the magnetic environment in which such eruptions occur allows one to get deeper insights into the relationship between linked eruptions.

\begin{figure}[ht]
\epsscale{1.0}
\plotone{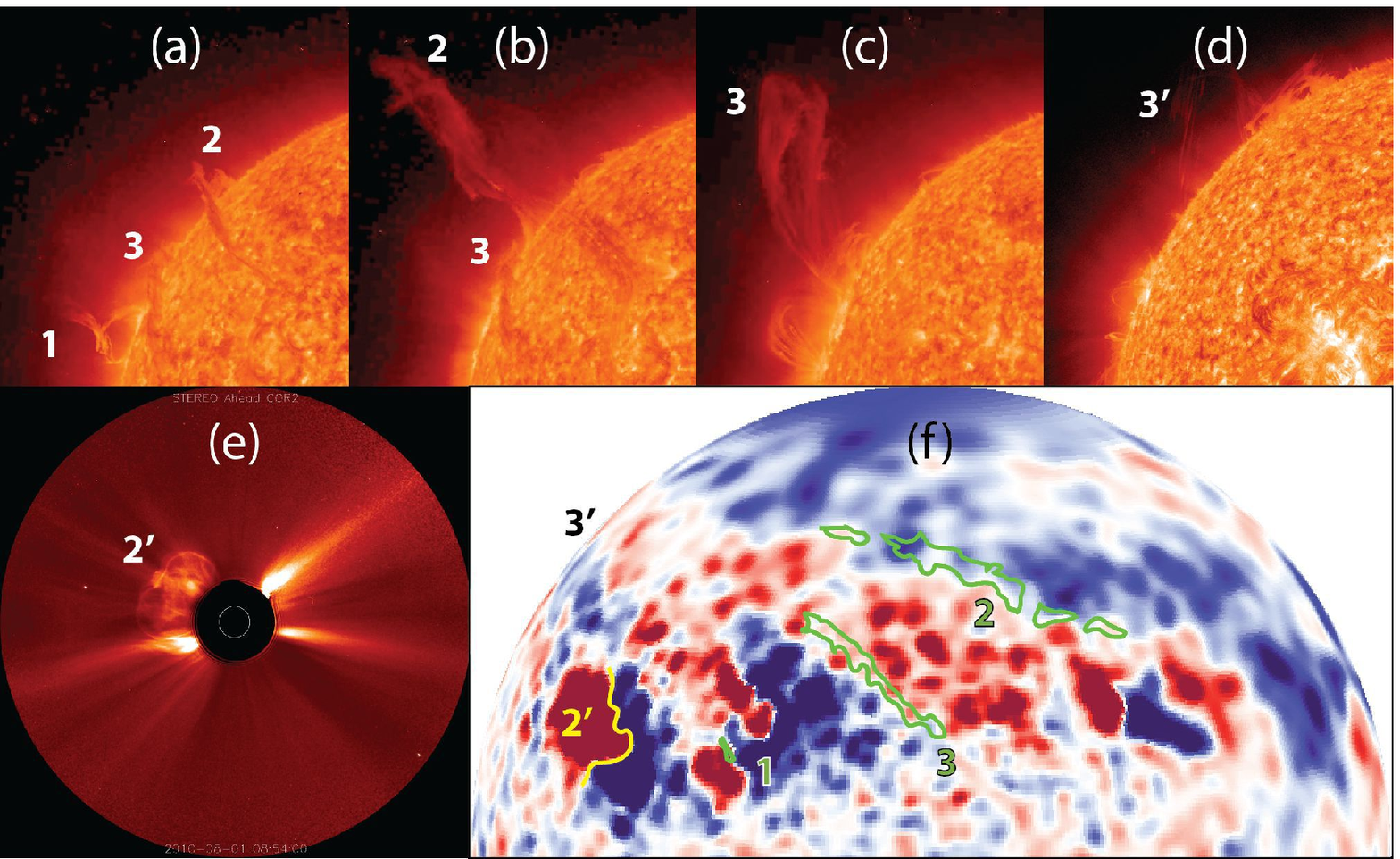}\\
\caption{
Sympathetic CMEs on 2010 August 1 with the main eruptions numbered in the order of their occurrence, primed numbers indicate near-simultaneous events; (a)-(c): eruptions 1, 2, and 3 as seen by STEREO-A 304 \AA\  at 02:56, 09:16, and 22:06 UT (left to right);  (d): eruption $3^{\prime}$ observed by SDO/AIA 304 \AA\  at 21:30 UT; (e): eruption $2^{\prime}$ captured by the COR2 coronagraph on board STEREO-A at 08:54 UT; (f): synoptic MDI magnetogram and contours (green) of the pre-eruption filaments that were visible in H$\alpha$, the yellow line indicates the location of the active-region filament $2^{\prime}$ prior eruption.    
	\label{f:symp_erups} }
\end{figure}

Figure\,\ref{f:symp_erups} shows that the sequence of eruptions started with a CME following the eruption of the small filament 1. 
About six hours later, the large quiescent filament 2 erupted, 
almost simultaneously with {\hM a C-class} flare and fast CME originating  {\hM in active region NOAA 11092} (whose polarity inversion line is denoted by $2^\prime$) to the east of filament 1. 
After another 12 hours, the  {\hM large} quiescent filament 3 erupted, again almost simultaneously with 
a large filament eruption (denoted by $3^\prime$)  that was observed above the eastern limb.
All of the large filament eruptions evolved into separate CMEs. 
{\hM Interestingly, while a filament was present along $2^\prime$, it did not erupt as part of the CME \citep{Liu2010b}.}

Our topological analysis of the large-scale background coronal field, which we
describe in detail in Section\,\ref{s:magtop}, reveals that, first, {\hM all of the erupting filaments} were located prior to their eruption below so-called {\hO pseudo-streamers} \citep[e.g.,][]{Hundhausen1972b, Wang2007b}. 
A pseudo-streamer is morphologically similar to a helmet streamer but, in contrast to it, divides coronal holes 
of the same rather than opposite polarity and contains two lobes of closed magnetic flux below its cusp to produce a {\hO  \PS-type structure}. These structures are quite common in the corona \citep[e.g.,][]{Eselevich1999} and are often 
observed to harbor filaments in their lobes \citep{Panasenco2010}. As the latter authors pointed 
out, an eruption in one lobe of a pseudo-streamer is often followed by 
an eruption in the other lobe shortly thereafter, indicating that these structures are prone to producing linked 
eruptions. 

Second, as suggested by \citet{Torok2011},  
the eruptions 2 and 3, which originated below one pseudo-streamer, were apparently 
triggered by eruption 1 that occurred outside the pseudo-streamer.
Third, as also suggested in that study, the fact that filament 2 erupted before filament 3, although it was located further from eruption 1 than filament 3, can be explained by the topological properties of the pseudo-streamer.

These three conclusions are strongly supported by our analysis in Sections\,\ref{s:magtop} and \ref{s:caus} and indicate the central role that pseudo-streamers may play in many linked eruptions.
We further develop this concept and generalize it in Section \ref{s:caus}, arguing that the order of {\em all} our eruptions, including those of filaments $2^{\prime}$ and $3^{\prime}$, is not coincidental but causal: It is essentially predetermined by the overall magnetic topology of the ensemble of pseudo-streamers that were involved into the eruptions.
We comprehensively investigate this topology in the framework of the potential field source-surface \citep[PFSS,][]{Altschuler1969, Schatten1969} model (Section \ref{s:config}), using new techniques for the structural analysis of magnetic fields (Section \ref{s:meth}).

Being of a general character, our findings on magnetic topology of pseudo-streamers have a broader impact than  was initially anticipated for this study.
In particular, they also provide important implications for the problem of the origin of the slow solar wind, which was recently addressed in the framework of the so-called S-web model \citep{Antiochos2007, Antiochos2011, Linker2011} and whose aspects have already been discussed in a number of papers \citep{Titov2011, Wang2012, Crooker2012}.
We address the implications of our new results for the S-web model in Section \ref{s:sweb} and summarize our work in Section \ref{s:s}.

{\hM

Although solar magnetic fields obtained from PFSS and MHD models often qualitatively match each other, at least if the latter are based only on line-of-sight magnetograms \citep{Riley2006}, it remains an open question whether the magnetic field topology, as understood in mathematical terms, is in both cases the same.
Section \ref{s:topSCs} makes it clear that this question indeed requires a special study, which is already on the way and will be described in the part II of a series of papers. 
In that part, we will repeat our analysis of the magnetic structure for the global solar MHD model derived from the same magnetogram as used in the present PFSS model. 
We also will compare the results of our analysis for both these models and, additionally, extend the discussion of these results, which we start in Section \ref{s:caus}, in relation to observations.

}


\section{INVESTIGATION METHODS}

\subsection{PFSS model}
	\label{s:config}


As a boundary condition for our PFSS model, we used the magnetic data that were derived from a {\hO SOHO/MDI synoptic map} of the radial field $B_r$ for Carrington rotation 2099 (July 13---August 9, 2010) using the Level 1.8.2 calibration.  We processed the synoptic map, first, by interpolating it to a uniform latitude-longitude mesh with a resolution of $0.5\arcdeg$. The {\hO polar} magnetic field was fitted in the new map with a geometrical specification to reduce noise in the poorly observed polar regions.   Second, we {\hO smoothed} the resulting $B_r$ by applying a diffusion operator such that its nonuniform diffusion coefficient was smaller in the active region and larger everywhere else.
Finally, we interpolated the obtained $B_r$ distribution from the uniform grid to a {\hO nonuniform} one that has a higher and lesser resolution, respectively, inside and outside the eruptive region.   {\hM This region} is spread in longitude and latitude approximately from $45\arcdeg$ to $180\arcdeg$ and from $-20\arcdeg$ to $65\arcdeg$, respectively,
with the resolution ranging from $0.37\arcdeg\times 0.37\arcdeg$ in this region to $2.6\arcdeg\times 1.8\arcdeg$ outside (see Figure \ref{f:maps1}(a)).

The spherical source surface, at which the scalar magnetic potential is set to be constant, is chosen at $r=2.5 R_{\sun}$,  where $R_{\sun}$ is the solar radius.  
%
%
For such a PFSS model, we have computed the photospheric map of coronal holes on a uniform grid with an angular cell size of $0.125\arcdeg$, which is much smaller than the smallest grid cell for the computed field itself.
The result is shown in Figure \ref{f:maps1}(b) together with the source-surface distribution of the squashing factor $Q$, which will be discussed below.
The {\hO three coronal holes} of negative polarity that are located in the eruptive region are distinctly disconnected from each other and from the negative northern polar coronal hole.
As will become clear later, the presence of these coronal holes in the eruptive region is crucial for understanding both the underlying magnetic topology and the plausible casual link that this topology sets up between the erupting filaments.

\subsection{Techniques for analyzing magnetic structure}
	\label{s:meth}


Magnetic configurations can generally have both {\it separatrix surfaces}  and {\it QSLs}.
To comprehensively  {\hO analyze}  the structure of our configuration, it is necessary to determine all such structural features, whose complete set we call the {\it structural skeleton} of the configuration.
We fulfill this task in two steps: First, we identify the footprints of the corresponding (quasi)-separatrix surfaces at the photosphere and source surface by calculating the distributions of the {\hO \it squashing} {\hO \it factor} $Q$ of elemental magnetic flux tubes \citep{Titov2002,Titov2007a};
these footprints are simply high-$Q$ lines of the calculated distributions.
Second, using the found footprints as a guide, we trace a number of field lines that best represent these surfaces.
%


For the calculation of $Q$ we use its definition in {\hO spherical coordinates} \citep{Titov2007a,Titov2008a}.
By construction, the $Q$ factor has the {\hO \it  same} value at the conjugate footpoints, so it can be used as a {\hO marker} for field lines.
In other words, despite being originally defined at the boundary surfaces only, the $Q$ factor can be extended into the volume by simply transporting its defined values along the field lines according to the equation
\begin{eqnarray*}
   {\bf B \cdot \nabla} Q = 0 \,,
\end{eqnarray*}
{\hM where} ${\bf B}$ is a given coronal magnetic field and $Q$ is an unknown function of space coordinates.
This equation can be solved in many different ways depending on the desirable accuracy and efficiency of the computation.
We will describe our methods for extending $Q$ in the volume in a future article together with other techniques for investigating (quasi-)separatrix surfaces, while here we would like to outline a few relevant considerations.

The extension of $Q$ in the volume makes it possible to determine the structural skeleton as a set of high-$Q$ layers.
{\hM They can intersect each other in a rather complicated way, especially low in the corona.
With increasing height, however, the intersections become simpler, which particularly helps our goal of studying the large-scale structure.
Determining the $Q$ distribution at a given cut plane, similar as done before in other works \citep{Aulanier2005, Titov2008a, Pariat2012, Savcheva2012b, Savcheva2012a}, is also helpful for analyzing complex structures.
We calculate $Q$ distributions at cut planes, extending the method that \citet{Pariat2012} described for configurations with plane boundaries to the case of spherical boundaries.
}
The high-$Q$ lines in such distributions visualize the cuts of the structural skeleton by those planes.
As will be shown below (Figure \ref{f:cutPS1}), this kind of visualization becomes particularly useful if the colors corresponding to {\hO low values} of $Q$ ($\lesssim 10^2$) are chosen to be {\hO transparent}.

We also find it useful to apply this transparency technique to the photospheric and source-surface $Q$ distributions, particularly if one uses in addition a special color coding that takes into account the local sign of the normal field  $B_{r}$ at the boundary.
The function that {\hM facilitates} this color coding is called {\hO \it  signed log Q} or simply $\slog Q$ and defined as \citep{Titov2011}
\begin{eqnarray}
  \slog Q \equiv \sign(B_r) \log\left[Q/2 + \left(Q^2/4-1\right)^{1/2}\right] \,.
	\label{slogQ}
\end{eqnarray}
Using a symmetric blue-white-red palette in combination with the above transparency mask, {\hM we make visible in $\slog Q$ distributions only high-$Q$ lines}, colored either in blue or red in negative or positive polarities, respectively. 
The resulting maps provide a compact and powerful representation of the structural skeleton at the boundaries, as evident from our illustrations below.

Since our magnetic field is potential, $Q$ acquires high values only in three cases: Either the corresponding field lines {\hO scatter} from localized inhomogeneities of the field nearby {\hO its null points or minimum points} \citep{Titov2009}, or touch so-called {\hO ``bald patches"} (BPs), which are certain segments of the photospheric polarity inversion line  \citep{Seehafer1986a, Titov1993}.
To make the whole analysis comprehensive, we separately determine the location of all such relevant features and then relate them to the high-$Q$ lines at the boundaries by tracing a number of field lines that pass through these features.
{\hM
The pattern of high-$Q$ lines determined at spherical surfaces of different radii provides us with estimates of the regions in which the  magnetic nulls and minima can be present.
Using then standard numeric algorithms \citep[see, e.g.,][]{Press2007}, both these features are found as local minima of $B^2$ that is defined between the grid points in these regions by cubic spline interpolation.
Calculation of the matrix of magnetic field gradients $\left[\nabla {\bf B}\right]$ and its eigenvectors at the found nulls and minima allows us to determine the local (quasi-)separatrix structure, which is further used to initialize tracing of the respective (quasi-)separatrix field lines.
For tracing generalized (quasi-)separators (see Section \ref{s:magtop}), which connect a pair of any of the above three features (i.e., nulls, minima, or BP points), we use a technique that is based on similar principles as described earlier for classical null-null separators by \citet{Close2004} and \citet{Haynes2010}.
}


\section{ANALYSIS OF THE MAGNETIC STRUCTURE}
	\label{s:magtop}

\subsection{Coronal holes versus high-$Q$ lines at the boundaries}
	\label{s:hQl}

As mentioned in Section \ref{s:config}, the eruptive region contains three coronal holes of negative polarity that are distinctly {\hO disconnected} at the photospheric level by positive parasitic polarities.
With increasing height, however, these coronal holes start to expand and subsequently merge with  each other and with the main body of the northern polar coronal hole.
Being fully open at the source-surface, the magnetic fluxes of these coronal holes still remain separated by the so-called {\hO \it separatrix curtains} \citep[SCs,][]{Titov2011}.
As will become clear below, the separatrix curtains are simply vertical separatrix surfaces that originate at null points of the magnetic field low in the corona.
At the source surface, their footprints appear as arcs joined at both ends to the null line of the magnetic field, so that the corresponding junction points divide the null line into several segments.
{\hM
Taken in different combinations, such segments and footprints of separatrix curtains form several closed contours.
The contours encompass the fluxes corresponding to the coronal holes that are disconnected at the photospheric level from each other and from the like-polarity coronal holes at the poles.
This fact clearly manifests itself on our source-surface $\slog Q$ map that is superimposed in Figure \ref{f:maps1}(b) on top of the photospheric coronal holes' map.
The figure indicates, in particular, that the high-$Q$ line of the footprint \SC{2} (\SC{3}) and the null-line segment to which the footprint adjoins encompass the \CH{2} (\CH{3}) flux.
Similarly, the source-surface footprints \SC{1} and \SC{2} and two short null-line segments to which the footprints adjoin encompass the \CH{1} flux.
}

It should be noted, however, that some of the source-surface high-$Q$ lines do not represent  the footprints of separatrix curtains,  but rather the footprints of  {\hO QSLs} that stem at the photosphere from narrow open-field corridors connecting spaced parts of otherwise single coronal holes.
The high-$Q$ lines of QSLs usually appear less sharp than those of separatrix-curtain footprints (see Figure \ref{f:maps1}(b)).
The indicated QSL footprints can easily be related to certain open field corridors in the northern polar coronal hole.
If one traces down several field lines from the paths that go across these high-$Q$ lines, the photospheric footpoints of these field lines will sweep along the respective open-field corridors, as predicted earlier by \citet{Antiochos2007}.
However, a similar procedure in the case of the separatrix curtains would give a very different result, which becomes clear after analyzing the magnetic topology low in the corona near the indicated coronal holes.

As a first step in this analysis, let us consider the coronal-hole maps and $\slog Q$ distribution, both defined at the photospheric level and superimposed onto each other as shown in Figure \ref{f:maps2}.
The pattern of high-$Q$ lines here is more complicated than at the source surface, as expected.
Nevertheless, in the region of interest, it prominently reveals three high-$Q$ lines (red), which are identified after inspection as photospheric footprints of the above-mentioned separatrix curtains.
They traverse along parasitic polarities and separate the indicated coronal holes in a similar manner as their source-surface counterparts.
Note also that these footprints and nearby filaments are locally co-aligned, and at least five of these filaments were eruptive.

\begin{figure}[]
\epsscale{1.0}
\plotone{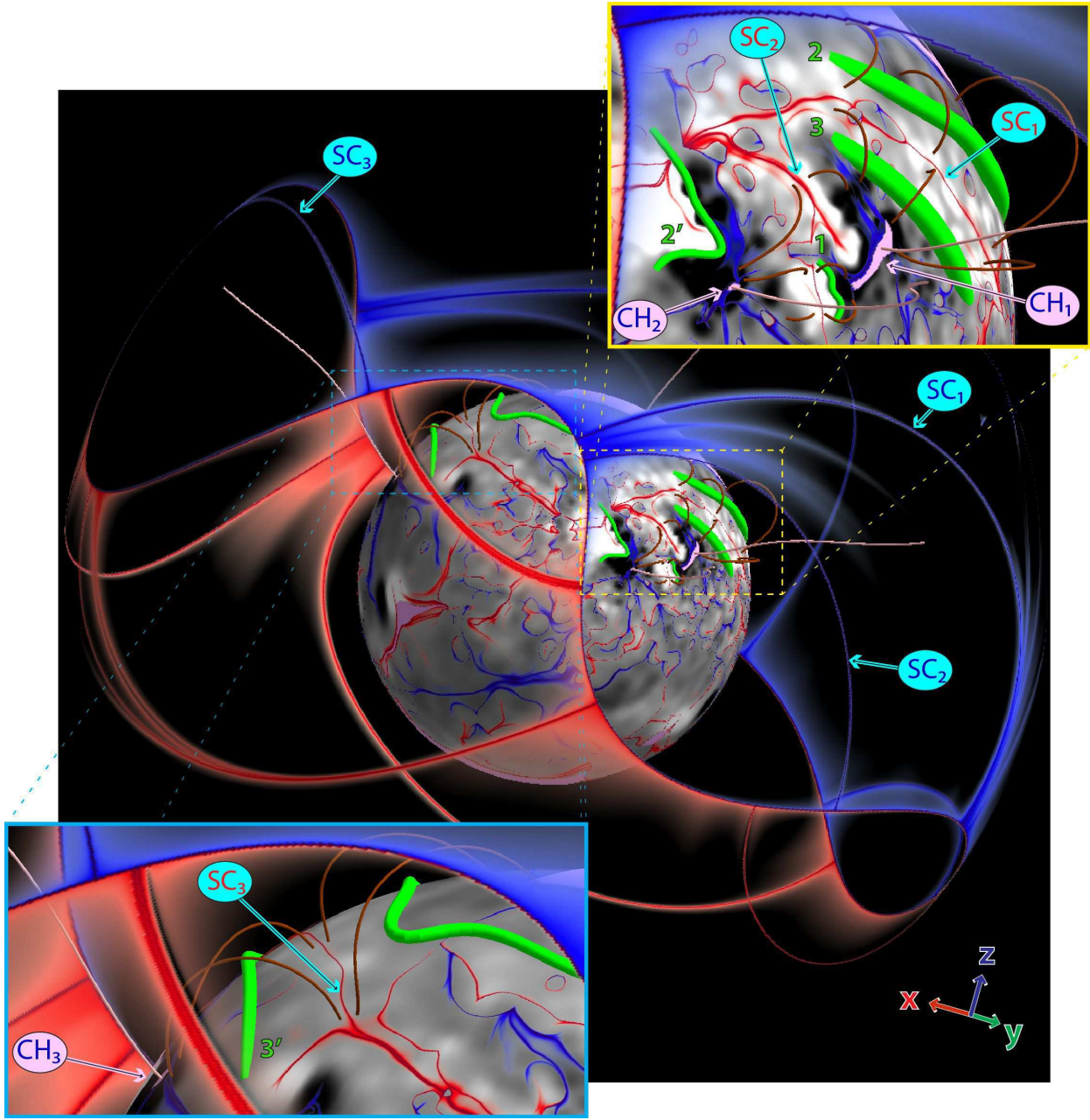}\\
\caption{$\slog Q$ distributions are mapped on the photospheric and source-surface globes with a varying opacity such that the low-$Q$ areas ($Q \lesssim 300$) appear to be fully transparent.  The photospheric $\slog Q$ map is superimposed on the respective grayscale $B_r$ distribution with the coronal holes shaded in light magenta.   Green tubes  depict the major filaments prior to the onset of sympathetic eruptions and several field lines (brown) indicate the pseudo-streamer lobes enclosing these filaments.   Open field lines (colored in pink) start in the middle of the coronal holes closest to the pseudo-streamers.  Vector triad in the lower right-hand corner indicates the angle orientation of the Cartesian system that is rigidly bound to the Sun center with the $z$-axis directed to the north pole.
	\label{f:s-web} }
\end{figure}

%
Figure \ref{f:s-web} shows the described distributions of $\slog Q$ and $B_{r}$ in three dimensions and a few field lines that produce loop-arcade structures above the filaments.
The loops of arcades are rooted with one footpoint at the positive parasitic polarities that disconnect our three coronal holes either from each other (\CH{1} from \CH{2}) or (\CH{1} and \CH{3}) from the northern coronal hole.
Thus, these arcades form in pairs the {\hO twin magnetic field lobes} of the three pseudo-streamers embedded between the indicated coronal holes. 
We also see here that four of the five filaments (all the numbered ones, except for $2^{\prime}$, in Figures \ref{f:maps2} and \ref{f:s-web}) were initially located inside such lobes.

\subsection{Separatrix structure of pseudo-streamers}

Of particular interest to us is the question on how the pseudo-streamer lobes are bounded in our configuration by separatrix surfaces of the magnetic field.
It turns out that these surfaces originate either at the null points or at the bald patches, both  mentioned already in Section \ref{s:meth} in connection with high-$Q$ lines.
Following \citet{Priest1996}, we will use the terms "fan surface" and "spine line" to designate, respectively, two-dimensional and one-dimensional separatrix structures that are related to a null point.
 They are defined through the eigenvectors of the matrix of magnetic field gradients at this point in the following way.
The fan separatrix surface is woven from the field lines that start at the null point in the plane spanned on the eigenvectors, whose eigenvalues are of the same sign.
The spine line is a separatrix field line that reaches the null point along the remaining third eigenvector.
For a potential field, the spine line is always perpendicular to the fan surface.

In accordance with the recent analytical model of pseudo-streamers \citep{Titov2011}, the boundaries of our pseudo-streamers are composed of three types of separatrix surfaces, two of which are the fan surfaces of some coronal null points, while the third one is a BP separatrix surface.
The fans of the first type have a curtain-like shape, whose field lines emanate from a null point, called henceforth basic one.
We have already discussed these surfaces above as separatrix curtains in connection with boundary high-$Q$ lines.
They contain both closed and open field lines and extend from the photosphere to the source surface, as shown in Figures \ref{f:SC1} and \ref{f:SC2}.

The fans of the second type are associated with other nulls and include only closed field lines.
Each of these fans 
bounds the closed flux of the parasitic polarity in a given pseudo-streamer only at one flank and forms a half-dome-like surface, whose edge is located in the middle of the pseudo-streamer and coincides exactly with the spine line of the basic null point  (see Figure \ref{f:dom12}).
The second half-dome is formed in all our three pseudo-streamers by the third type of separatrix surfaces that originate in BPs at the opposite flanks of pseudo-streamers.
In fact, in our third pseudo-streamer even both separatrix half-domes are due to the presence of BPs (Figure \ref{f:SC3}).

\subsection{Field line topology of separatrix curtains}  \label{s:topSCs}

\begin{figure}[]
\epsscale{1.0}
\plotone{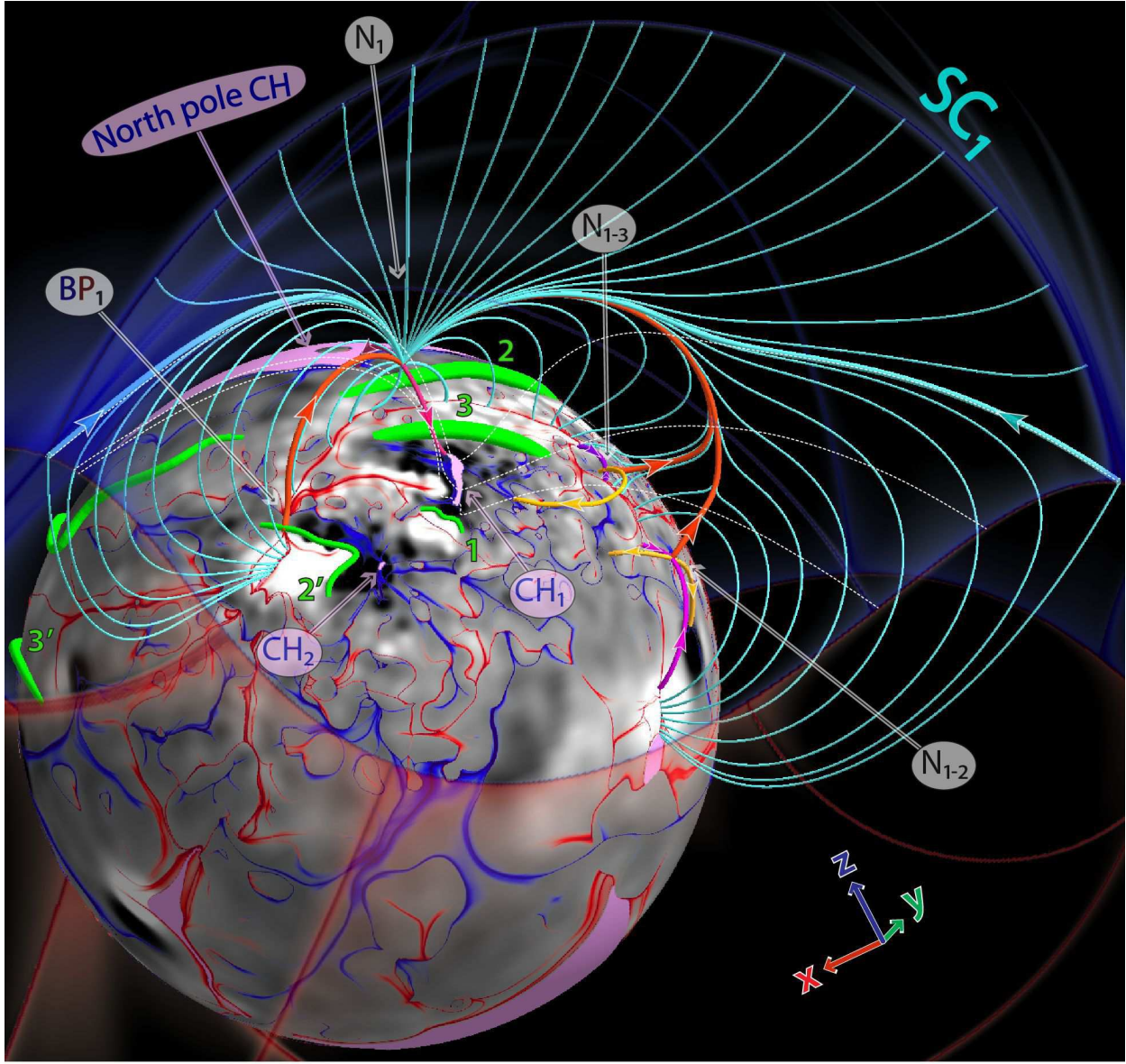}\\
\caption{Field line topology of the separatrix curtain \SC{1} of pseudo-streamer 1 (embedded between the northern polar coronal hole and  \CH{1}).
The thickest lines represent separators, of which the red ones are closed field lines connecting the null point \N{1} either to the bald patch \BP{1} or another nulls \N{1\mbox{-}2} or \N{1\mbox{-}3}, while the cyan ones are open field lines connecting \N{1} to the null line of the source surface.
Magenta lines are the spine field lines of the nulls; the yellow lines are the separatrix field lines that emanate from the nulls \N{1\mbox{-}2} and \N{1\mbox{-}3} along the fan eigenvectors that are complementary to the separator ones; several field lines (white dashed) belonging to the boundary of \CH{1} are also shown.
The maps at the photosphere and source surface and their color coding are the same as in Figure \ref{f:s-web}.
	\label{f:SC1} }
\end{figure}

\begin{figure}[]
\epsscale{1.0}
\plotone{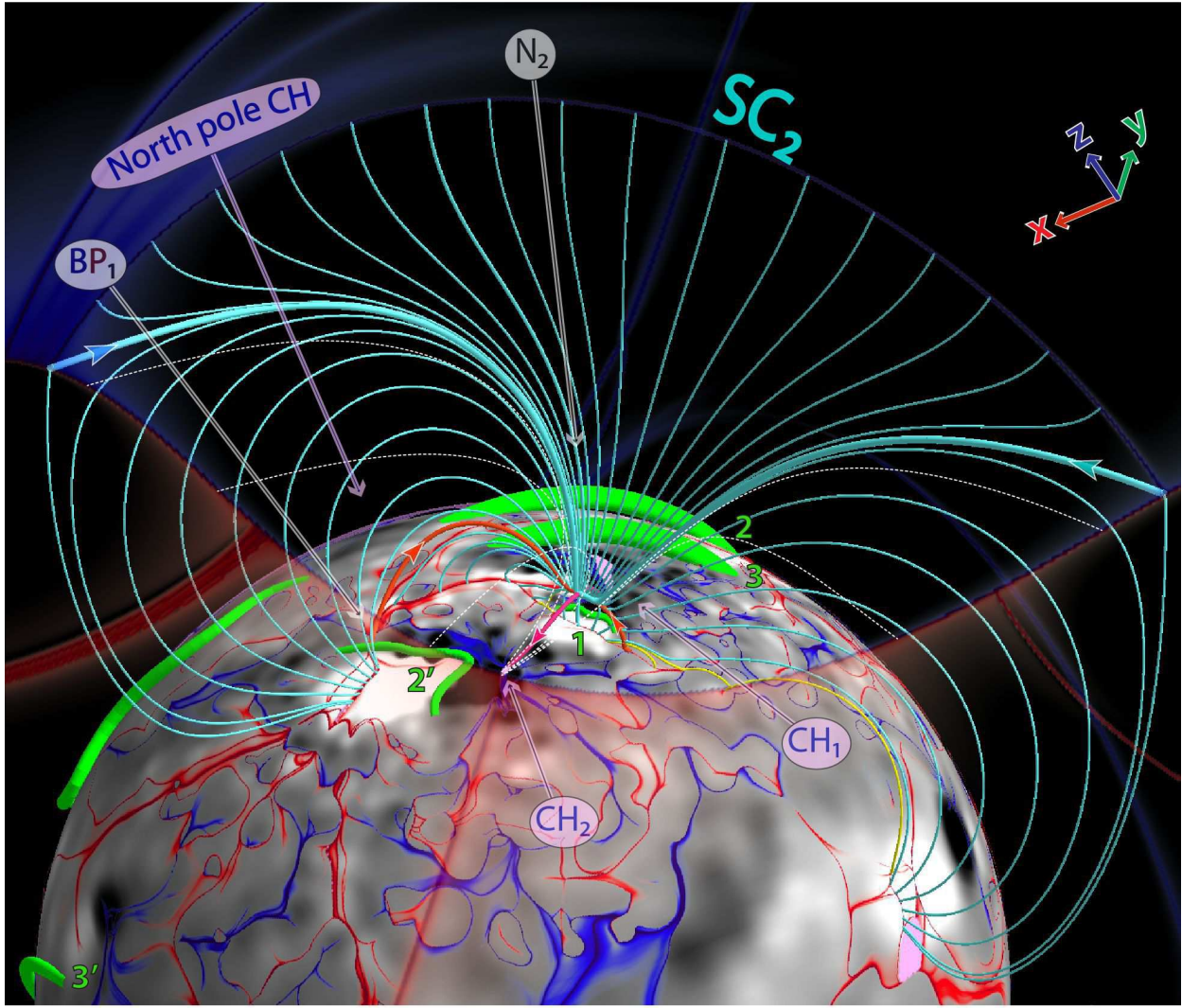}\\
\caption{Field line topology of the separatrix curtain \SC{2} of pseudo-streamer 2 (embedded between the coronal holes \CH{1} and \CH{2}).
The field line styles are the same as in Figure \ref{f:SC1}, except that the thin yellow lines represent separatrix field lines associated with small-scale photospheric polarity regions. 
The maps at the photosphere and source surface and their color coding are the same as in Figure \ref{f:s-web}.
	\label{f:SC2} }
\end{figure}

\begin{figure}[]
\epsscale{1.0}
\plotone{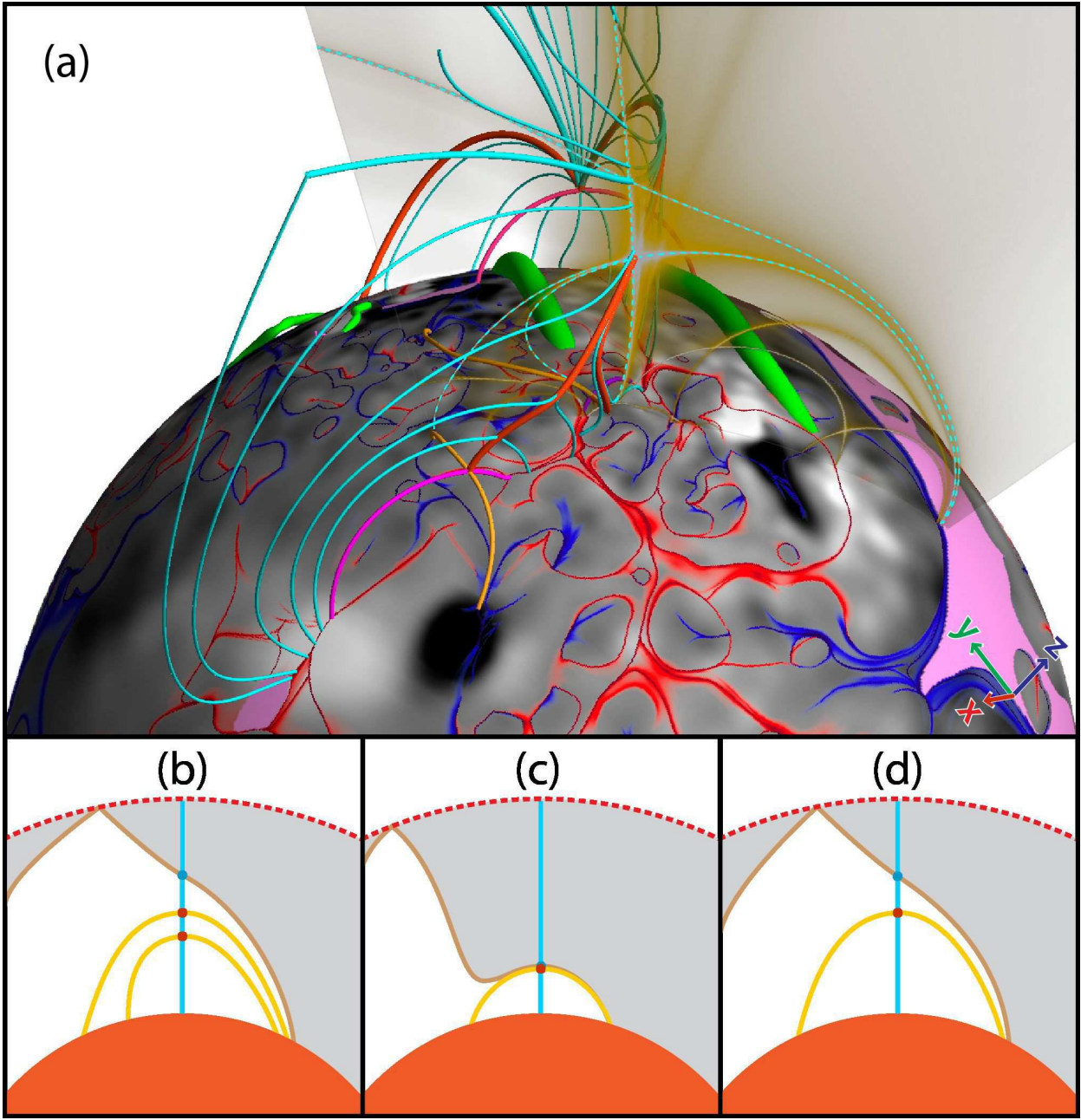}\\
\caption{Field line topology of the separatrix curtain \SC{1} in relation to the $\log Q$ distribution in a cut plane across the pseudo-streamer 1 (a).
This distribution is plotted by using a yellow palette, whose opacity linearly decreases with $\log Q$ in the range from 2.5 to 0.3 down to a complete transparency;
the maps at the photosphere, their color coding, and the field line styles are the same as in Figure \ref{f:SC1}.
Dashed (cyan) curves highlight the high-Q lines that represent the intersection lines of the cut plane with \SC{1}, helmet-streamer separatrix surface, and two separatrix domes.
Such a structure is shown also schematically for this cut and two others in panels (b), (c), and (d), respectively, where the open field regions are shaded in gray; the extra two cuts are made successively further eastward from the middle of the pseudo-streamer.
	\label{f:cutPS1} }
\end{figure}

Consider now in more detail the field line topology in all our three pseudo-streamers, starting from the two neighboring separatrix curtains \SC{1} and \SC{2} (see Figures \ref{f:SC1} and \ref{f:SC2}).
The field lines in each of these curtains fan out from its own basic null point that is located between two adjacent coronal holes of like polarity and above the respective parasitic polarity.
The footprints of separatrix curtains, which are discussed in Section \ref{s:hQl}, can be viewed then as photospheric or source-surface images of single null points \N{1} and \N{2}  due to their mapping along closed or open, respectively, field lines.

Within a given separatrix curtain, such a mapping is continuous everywhere except for few special field lines, called {\hO \it separators}, where the mapping suffers a jump.
This jump takes place whenever a mapping field line hits a null point (like \N{1\mbox{-}2} and \N{1\mbox{-}3} in Figure \ref{f:SC1}, or a bald patch, like \BP{1} in both Figures  \ref{f:SC1} and  \ref{f:SC2}).
To distinguish these separators from other field lines, we have plotted them thicker in these and further similar figures.

In addition to the mentioned closed separators, there are also two open ones for each of the curtains.
These open separators connect the null \N{1} (or \N{2}) to a pair of null points belonging to the source-surface null line.
The latter is simply the helmet streamer cusp, from which the heliospheric current sheet arises.
Each of these pairs of nulls also coincide with the end points of the source-surface footprints of separatrix curtains.

Note, however, that any null line of the magnetic field is a topologically unstable feature that can exist only under very special conditions.
We think, therefore, that the source-surface null line is most likely an artifact inherent only in the employed PFSS model.
If passing from PFSS to MHD model, such a null line must turn at radii close to $2.5 R_{\sun}$ into a feature that has a substantially different magnetic topology.
Thus, the indicated topological linkages have yet to be refined, using a more realistic than PFSS model of the solar corona.
We will do that in {\hM our next  paper II}, while here we proceed the analysis, assuming that our findings on open separators are approximately correct.

%

\subsection{Field line topology of separatrix half-domes}

\begin{figure}[]
\epsscale{1.0}
\plotone{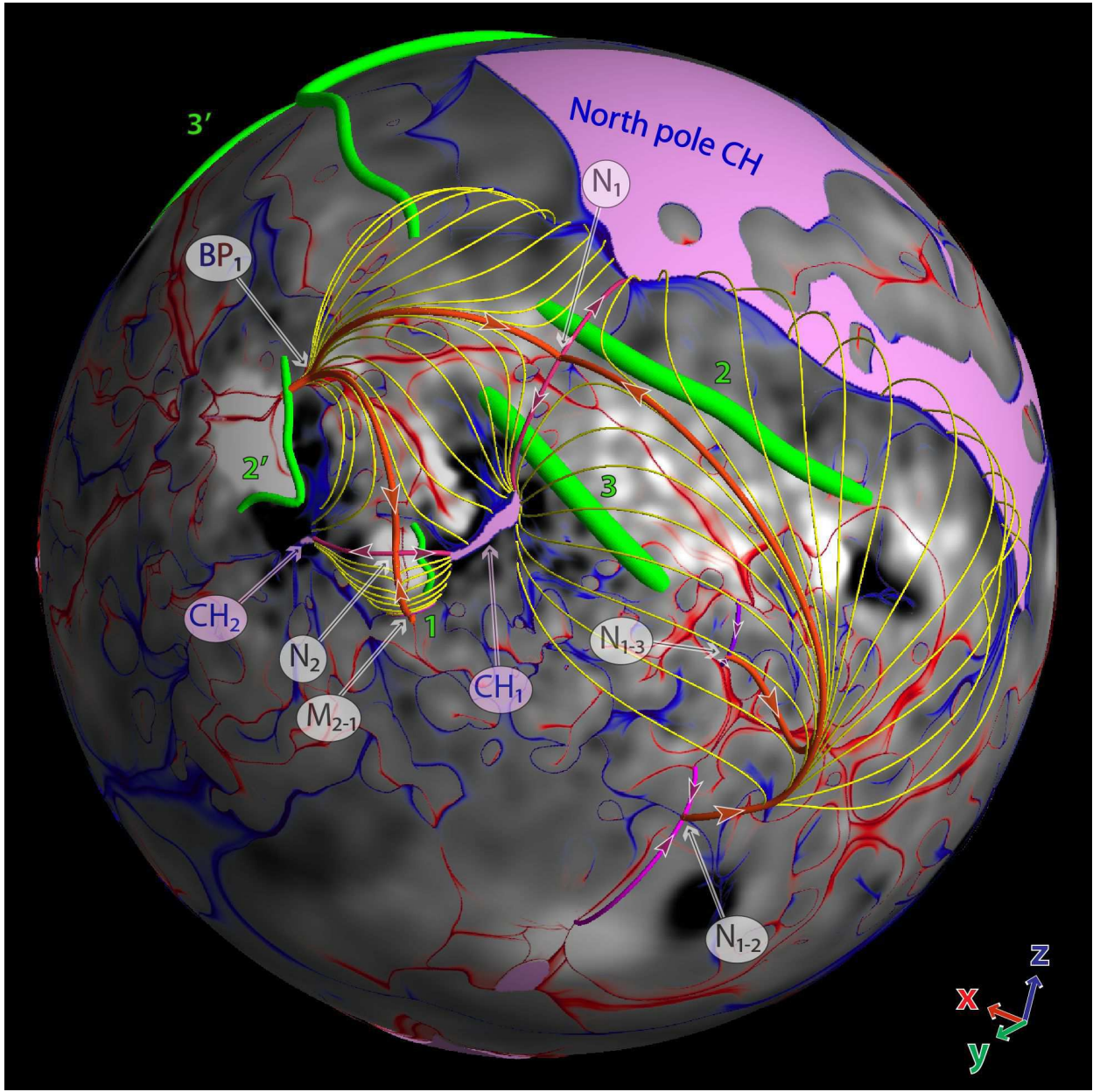}\\
\caption{Field line topology of the separatrix domes of pseudo-streamers 1 and 2, one of which is embedded between the northern polar coronal hole and  \CH{1}  and the other between the coronal holes \CH{1} and \CH{2}.
The field line styles are the same as in Figures \ref{f:SC1} and \ref{f:SC2}, except that the thin yellow lines represent separatrix field lines starting either at the bald patch \BP{1} or in the fan plane of the null point  \N{1\mbox{-}2}; a similar separatrix dome associated with the null \N{1\mbox{-}3} is not shown.
The same style is used for the field lines of the quasi-separatrix surface originated at the magnetic minimum point \M{2\mbox{-}1}.
The maps at the photosphere and source surface and their color coding are the same as in Figures \ref{f:s-web}--\ref{f:SC2}.
	\label{f:dom12} }
\end{figure}

\begin{figure}[]
\epsscale{1.0}
\plotone{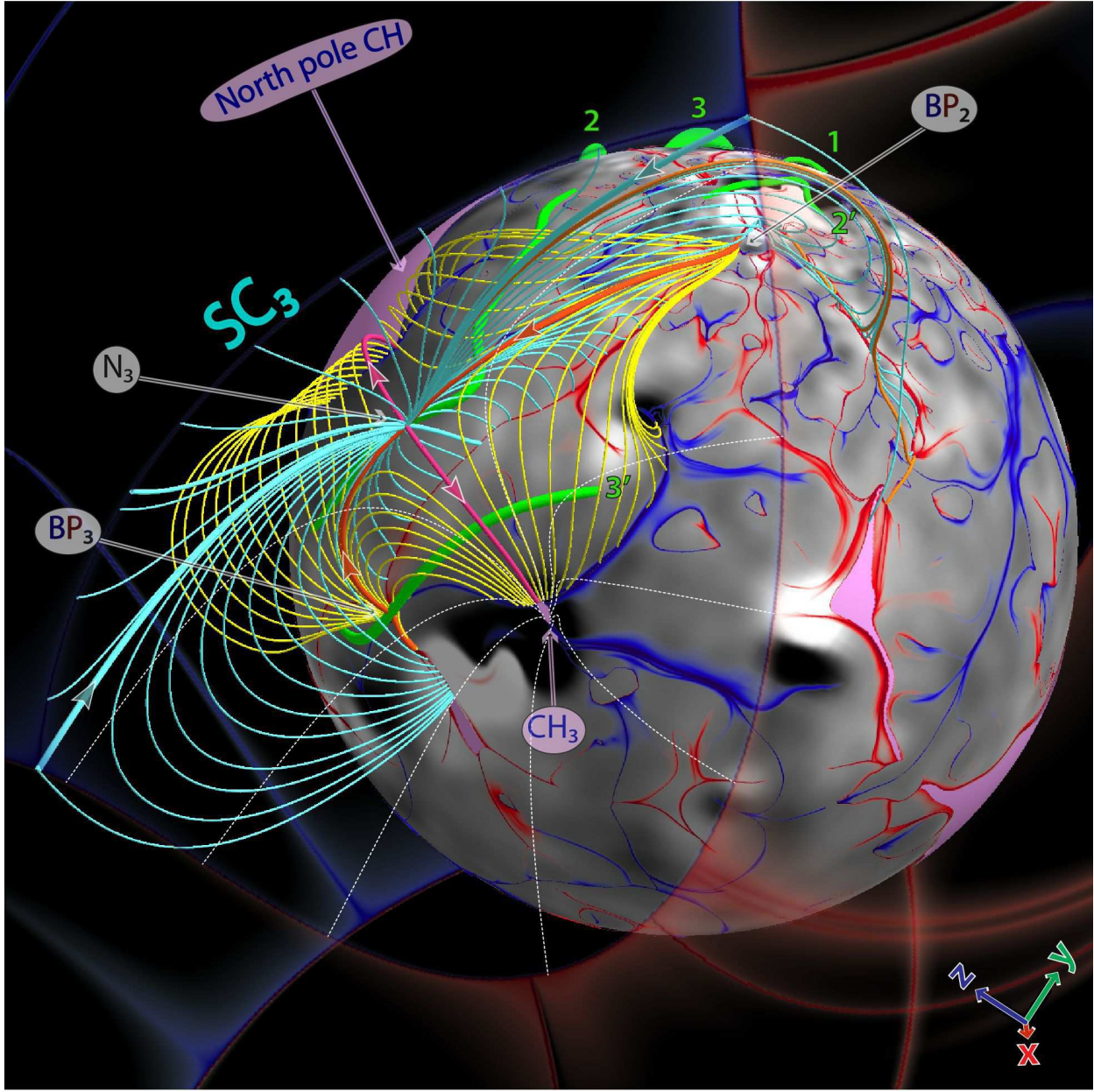}\\
\caption{Field line topology of the separatrix curtain (cyan) and dome (yellow) of pseudo-streamer 3 (embedded between the northern polar CH and  \CH{3}).
The field line styles are the same as in Figures \ref{f:SC1}--\ref{f:dom12}, except that the thin orange lines represent the separatrix field lines that are associated with the bald patches and null points of small-scale photospheric polarity regions.
The maps at the photosphere and source surface and their color coding are the same as in Figures \ref{f:s-web}--\ref{f:dom12}.
	\label{f:SC3} }
\end{figure}

Consider now in more detail the topology of separatrix domes (Figure \ref{f:dom12}), starting from the pseudo-streamers that are embedded between \CH{1}, \CH{2}, and the northern polar coronal hole.
The eastern half-domes (on the left) are combined in one simply connected surface, because they originate in one small bald patch \BP{1} located at the border of an active region near the filament $2^{\prime}$.
Spreading out from \BP{1}, the field lines extremely diverge within this surface at the nulls \N{1} and \N{2} and hit the photosphere near the indicated coronal holes.
Two of these field lines (red and thick), however, go instead straight to \N{1} and \N{2} and so, as discussed above, are {\hO \it generalized separators} belonging to \SC{1} and \SC{2}, respectively.

In contrast to the eastern half-domes, the western ones (on the right) do not merge with each other and have different originations.
The half-dome covering filaments 2 and 3 is simply a fan surface of an extra null point \N{1\mbox{-}2} that is located far to the west from the basic null \N{1}.
These two nulls are connected by an ordinary separator, which belongs to both this half-dome and the curtain \SC{1}.

It is a bit surprising, but
the half-dome covering filament 1 appears to be a quasi-separatrix surface that originates at a magnetic minimum point \M{2\mbox{-}1} lying very close to the photosphere.
The field line (red and thick) that passes through and connect \M{2\mbox{-}1} to the basic null \N{2} is a {\hO \it quasi-separator\/}.
The field direction remains unchanged after passing this line through the minimum \M{2\mbox{-}1}, as opposed to a genuine null point, where the field direction would change to the opposite.
A similar behavior of the field at \M{2\mbox{-}1} would also occur if it were a degenerate null point, whose one eigenvalue identically equals zero \citep{Titov2011}.
We regard this possibility as highly unlikely here, but we cannot either fully exclude it, relying only on our numerical study as approximate of nature.

%
%
The existence of the null \N{1\mbox{-}3} in the first of the two discussed pseudo-streamers brings an extra complexity into the structure.
Figures \ref{f:SC1} and \ref{f:dom12} show that, similarly to \N{1\mbox{-}2}, the null \N{1\mbox{-}3} is  connected via an additional separator to the basic null \N{1}.
This implies that the fan surface of \N{1\mbox{-}3} is also a half-dome such that its edge coincides with the spine line of the null \N{1}.
We did not plot this half-dome in Figure \ref{f:dom12} to avoid cluttering the image with too many lines, but it is very similar to the plotted half-dome that originates in the null \N{1\mbox{-}2}.

The third pseudo-streamer, which is embedded between \CH{3} and the northern polar coronal hole, has the  topology as analogous as the one of the two others considered above (see Figure \ref{f:SC3}).
The main difference is only that both half-domes originate here at bald patches \BP{2} and \BP{3}, which are located at the opposite flanks of the pseudo-streamer.
In this respect, the structure is the same as the one used before for initializing our MHD model of sympathetic eruptions \citep{Torok2011}.
It is important that these simulations have demonstrated that the generalized separators connecting such BPs and null points are physically similar to the ordinary separators.
They both appear to be preferred sites for the formation of current sheets and reconnection of magnetic fluxes.

\subsection{Field line topology versus high-$Q$ lines in the cut planes}

A complementary way to study the structure of a pseudo-streamer is to consider its cross-sectional $Q$ distributions and analyze their variation in response to changing location of the cut plane.
As one can anticipate from the above analysis, the simplest pattern of high-$Q$ lines appears to occur in the cut plane across the very middle of pseudo-streamers, where the basic null point is located.
The corresponding high-$Q$ lines form there a \PS-type intersection such that the vertical line and arc in the symbol \PS\ represent, respectively, the discussed separatrix curtains and domes.
The shape of separatrix domes at this place essentially follows the path of the spine line associated with the respective basic null point.
Above such a dome, the separatrix curtain separates the open fields of two adjacent coronal holes and observationally corresponds to the stalk of the pseudo-streamer.

However, with shifting the cut plane from the middle to the flanks of pseudo-streamers, the pattern of high-$Q$ lines gets more complicated.
In particular, the above high-$Q$ arc can split into several lines, each of which corresponds to a separate half-dome, except for the uppermost line.
The latter asymmetrically rises on one side from the curtain up to the source surface and, touching it, forms a cusp.
This line determines the border between closed and open fields, since it is nothing else than an intersection line of the cut plane with the separatrix surface of the helmet streamer.
Figure \ref{f:cutPS1} illustrates such a structure in a particular cut plane; it also shows schematically how the cross-sectional pattern varies along the pseudo-streamer.
Only three cases where the cut plane passes at the photospheric level outside \CH{1} and \CH{2} are shown in this figure, while the remaining cases can be reproduced analogously from the above analysis.

\subsection{Concluding remarks}

So far, we have fully described only the structural skeleton of the first pseudo-streamer, including the separatrix curtain \SC{1} and respectve half-domes with their separators.
As concerned with the other two pseudo-streamers, we still have not touched on several separators depicted in Figures \ref{f:SC2} and \ref{f:SC3} with yellow and orange thin lines.
These separators are due to ``scattering" of the separatrix-curtain field lines on small photospheric flux concentrations of negative polarity.
Such scattering occurs at bald patches or null points to yield additional half-domes, whose edges coincide with the spine lines of the basic nulls \N{2} or \N{3}.
The existence of these features, however, can vary depending on the resolution and smoothing of the used magnetic data, so we ignore them in our study, focusing only on stable structural features that are due to large-scale properties of the configuration.

One has also to remember that the described structure might be distorted in reality by the field of filaments whenever they are  present inside pseudo-streamer lobes.
Note, however, that such filaments reside prior to eruption in the middle of the lobes along photospheric polarity inversion lines.
So possible intense currents of the filaments are located relatively far from the found separatrix domes and curtains and hence the contribution of such currents to the total field must be small at these places compared to the background potential field.
Therefore, 
we think that at large length scales our PFSS model is accurate enough to describe the structure of the real pseudo-streamers with the filaments inside the lobes.


\section{MAGNETIC TOPOLOGY AS A CAUSAL LINK IN SYMPATHETIC CMES}  \label{s:caus}

We have studied in Section \ref{s:magtop} how separatrix curtains and half-domes originate in a given pseudo-streamer at magnetic null points and/or bald patches and how they intersect each other along separator field lines.
These results are of importance for unveiling a causal link in the sequential eruption of filaments, in which the magnetic topology and reconnection likely played a key role.
Indeed, according to the present state of knowledge \citep{Priest2000}, a perturbation in the neighborhood of a separator line generally creates along it a current sheet, across which magnetic fluxes subsequently reconnect in an amount depending on the form and strength of the perturbation.
As demonstrated above, each of our pseudo-streamers contains several separators, all of which are connected to a basic null point.
A perturbation in its neighborhood is expected then to cause reconnection along each of these separators, resulting ultimately in a flux redistribution between adjacent topological regions.

It follows from our analysis that these regions are simply the volumes bounded by various parts of the separatrix curtain, half-domes, and separatrix surface of the helmet streamer.
Unfortunately, such a complex topological partition of the volume makes it difficult
to foresee all the details of the response of our pseudo-streamers to different MHD perturbations.
It is clear, however, that eventually such perturbations will change the magnetic fluxes in the lobes and consequently the stability conditions for the filaments within them.
The latter in turn can influence the order of eruption of the filaments, which was recently demonstrated in our simple MHD model of sympathetic eruptions \citep{Torok2011}.
In this model, a pseudo-streamer similar to the one that stems from the basic null \N{1} played a key role in guiding the eruptions of the magnetic flux ropes, analogous to our filaments 2 and 3.
Thus, our present topological analysis of the potential background field further substantiates the model.

Let us put now the results of that model into the context of our present analysis in order to explain the observed sequence of the 2010 August 1--2 CMEs.
For simplicity, we restrict our consideration to the reconnection processes that occur in the vicinity of the basic nulls of the pseudo-streamers, where we expect the greatest perturbation to occur during the onset of eruptions.
As shown above,  all separatrix half-domes merge there and form together with the separatrix curtain a simple \PS-type intersection.
Such a separatrix structure implies that, irrespective of the form of the external perturbation, the reconnection triggered there will be of the interchange type \citep[e.g.,][]{Fisk2005}.
It will exchange the fluxes between the lobes and coronal holes in such a way that the sum of the fluxes in both the two lobes and the two coronal holes remains unchanged.
In other words, the diagonally opposite lobes and coronal holes form conjugate pairs, so that the flux in one pair increases by the same amount that it decreases in the other pair. 

To facilitate further discussion, we label the pseudo-streamers by the numeric label of their basic null; similarly, we label the lobes by the label of their embedded filament.  
Note, first, that erupting filament 1 resides initially in pseudo-streamer 2, which is located south of pseudo-streamer 1 (see Figure\ \ref{f:dom12}).  Therefore, the rise of filament 1 perturbs the southern side of pseudo-streamer 1 and eventually triggers interchange reconnection between the fluxes of coronal hole \CH{1} and lobe 2.  This reconnection reduces the flux in lobe 2, thereby removing the field lines that overlie and stabilize filament 2, eventually causing it to erupt (i.e., the second eruption).  On the other hand, this same interchange reconnection causes the flux in lobe 3 to increase, adding field lines that overlie filament 3, thus further stabilizing it.  However, later in time, after erupting filament 2 has risen to a sufficient height, a vertical current sheet forms in its wake, providing a site for interchange reconnection between the fluxes of lobe 3 and the northern polar coronal hole.  This second reconnection eventually reduces the flux in lobe 3, removing field lines that overlie and stabilize filament 3, eventually causing filament 3 to erupt (i.e., the third eruption).

This scenario is consistent with that proposed for the sequential eruption of  {\hM filaments 1--3} in our idealized model \citep{Torok2011}.
There is one difference though: our present PFSS model reveals that filament 1 was also located  inside a pseudo-streamer, which is pseudo-streamer 2 in our notation.
The presence of this pseudo-streamer, however, merely facilitates the eruption of filament 1, because its overlying field becomes open at a very low height. 
So this new feature fits nicely with our earlier proposed mechanism.

\begin{figure}[ht]
\epsscale{1.0}
\plotone{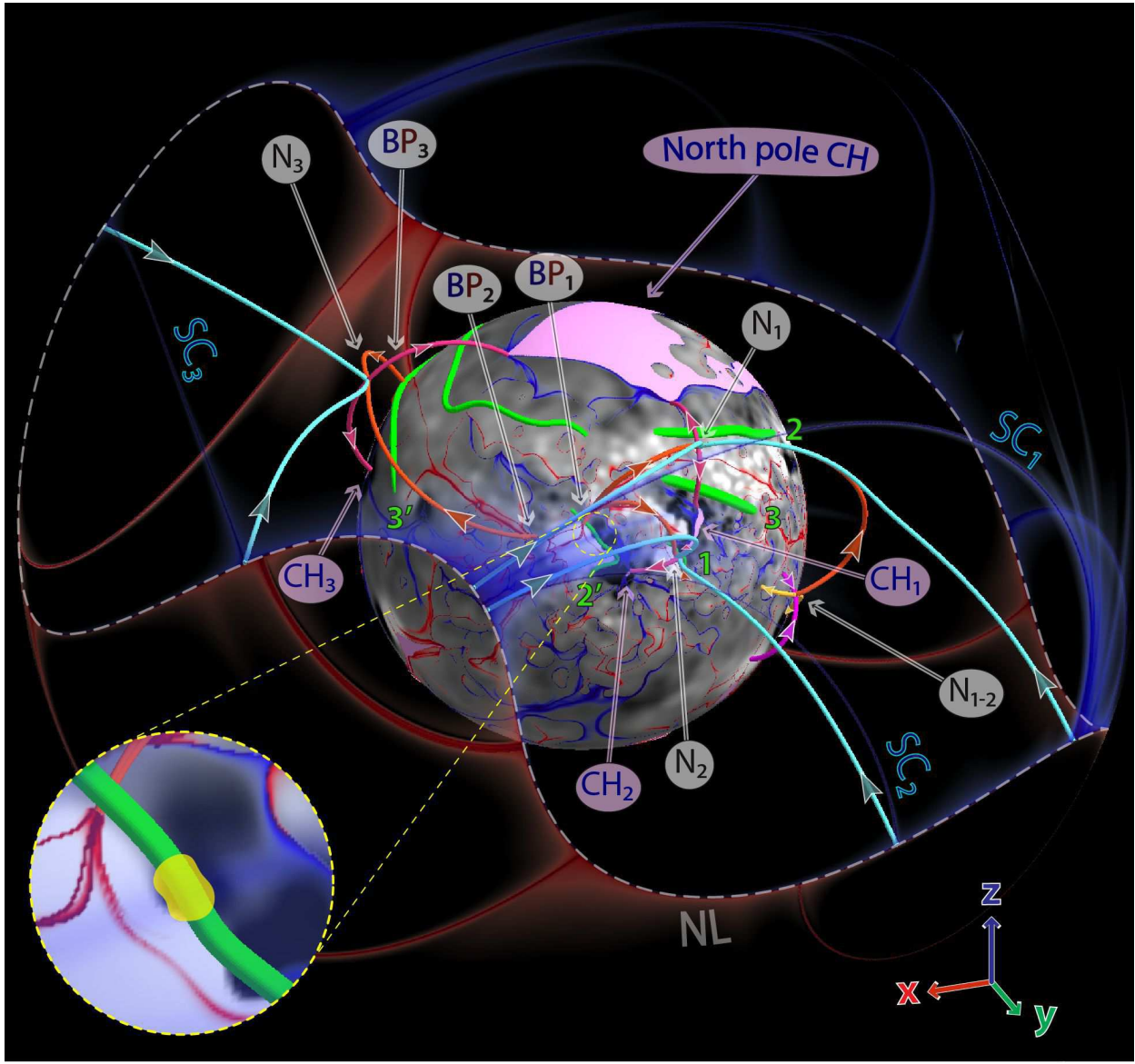}\\
\caption{The chain of separators and spine lines in all three pseudo-streamers that were involved in the 2010 Aug 1--2 sympathetic CMEs.
The white dashed line is the null magnetic field line of the source surface---together with the open separator field lines (cyan), it provides a global coupling between all three null points of the pseudo-streamer separatrix curtains.
The inset shows a zoomed region near \BP{1}, where a strong pre-flare brightening (indicated by yellow blob) was observed by SDO/AIA at $\sim$ 06:36 UT shortly after which eruption 2' started.
	\label{f:fuse} }
\end{figure}

The present analysis suggests  {\hM possible explanations} also for the eruptions $2^{\prime}$ and $3^{\prime}$.
{\hM According to} Figures \ref{f:SC1}--\ref{f:dom12}, filament $2^{\prime}$ passes above bald patch \BP{1}, which is connected by two separators to the basic null points \N{1} and \N{2}.
As discussed above, the rise of filaments 1 and 2 {\hM is expected to activate} these separators, forming current sheets along them, and subsequently triggering reconnection.
{\hR Around the location of \BP{1}, this reconnection may have been} 
of the tether-cutting type \citep{Moore2001}, reducing the confinement of {\hR the active region core field} and eventually unleashing its eruption. 
 This explanation is in agreement with the fact that SDO/AIA observed several brightenings in the active region before the {\hM CME occurred}.
There was a particularly strong brightening at $\sim$ 06:36 UT 
{\hM below and above filament $2^{\prime}$,}
very close to the bald patch \BP{1} (see the inset in Figure \ref{f:fuse}).
This brightening occurred after filament 2 had already started to rise, implying the above activation of the separator and subsequent reconnection {\hM in the vicinity of bald patch \BP{1}.} 
{\hM We note that \cite{Liu2010b} also associated the pre-eruption brightening at $\sim$ 06:36 UT to tether-cutting reconnection, triggered, however, by photospheric converging flows rather than separator activation.  It appears indeed possible that both process played a role.}
We will make a more detailed comparison of our topological analysis with observations {\hM in paper II}.

The location of pseudo-streamers 1 and 3 indicates that the 
{\hM eruptions} 2 and $2^{\prime}$ {\hM should produce} 
a significant perturbation of the northern side of pseudo-streamer 3.
This  {\hM should} lead to interchange reconnection between lobe $3^{\prime}$ and the northern polar coronal hole, reducing the magnetic flux in this lobe and eventually causing filament $3^{\prime}$ to erupt, in a similar way {\hM as} described for filament 2.  Notice also that filament $3^{\prime}$ rises above bald patch \BP{3}, which is connected by a separator
to  {\hM the} basic null \N{3} (see Figure \ref{f:SC3}).
{\hM As discussed above for eruption $2^{\prime}$, resulting tether-cutting reconnection may trigger the destabilization of filament $3^{\prime}$,}
in tandem with the indicated flux reduction in the lobe $3^{\prime}$ caused by interchange reconnection.

This concludes the extended scenario for the sympathetic eruptions under study.
Figure \ref{f:fuse} summarizes it, presenting all the topological features that are relevant for this scenario.
In particular, it depicts the closed separators (red thick lines) that form a long chain that traverses through all three pseudo-streamers.
As described above, such a separator chain likely sets up a global coupling between eruptions occurring at widely separated locations.
Figuratively speaking, this separator chain plays the role of a ``safety fuse" in which a single eruption at one end of the chain triggers along it a sequence of the observed electromagnetic explosions.

Additional global coupling between pseudo-streamers and {\hM eruptions} might also be provided by the open separators (thick cyan lines in Figure \ref{f:fuse}), which connect the basic nulls of the pseudo-streamers  to the cusp of the helmet streamer.
This coupling, however, has yet to be verified.  It requires a more advanced model than the PFSS model used in the present study.  We plan to use an MHD
model for this purpose in the next step of our study.

The proposed explanation of the assumed causal link in the observed sympathetic eruptions is of substantial heuristic value.
It is particularly useful as a guide for setting up and analyzing further numerical studies of these eruptions.
In combination with our structural analysis, more detailed numerical simulations of CMEs in this configuration are needed to prove the existence of such a link and to deepen its understanding.


\section{IMPLICATIONS FOR THE S-WEB MODEL}
	\label{s:sweb}

The structural analysis of pseudo-streamers that we have described has importnant implications not only for sympathetic CMEs but also for the slow solar wind.
The recent S-web model \citep{Antiochos2011, Linker2011} has sparked substantial interest in the community \citep{Crooker2012, Wang2012}.
Unfortunately, several important issues related to this model are not well understood.  Since the results obtained above relate to the S-web model, we will use this opportunity to clarify these issues.

The first issue relates to the concept of coronal hole connectivity.
Some confusion has arisen because the connectivity of coronal holes has been interpreted in two different senses. 
We can consider coronal holes either as two-dimensional regions at the photosphere or as three-dimensional regions in the corona.
Though coronal holes of like polarity are always connected 
when considered 
as three-dimensional regions,
it is important to note that they {\it can be disconnected} in the photosphere 
when considered 
as two-dimensional regions \citep{Titov2011}.  
In this case, they 
merge at some height in the corona via a field-line separatrix structure that observationally manifests itself as a pseudo-streamer.

\begin{figure}[]
\epsscale{0.85}
\plotone{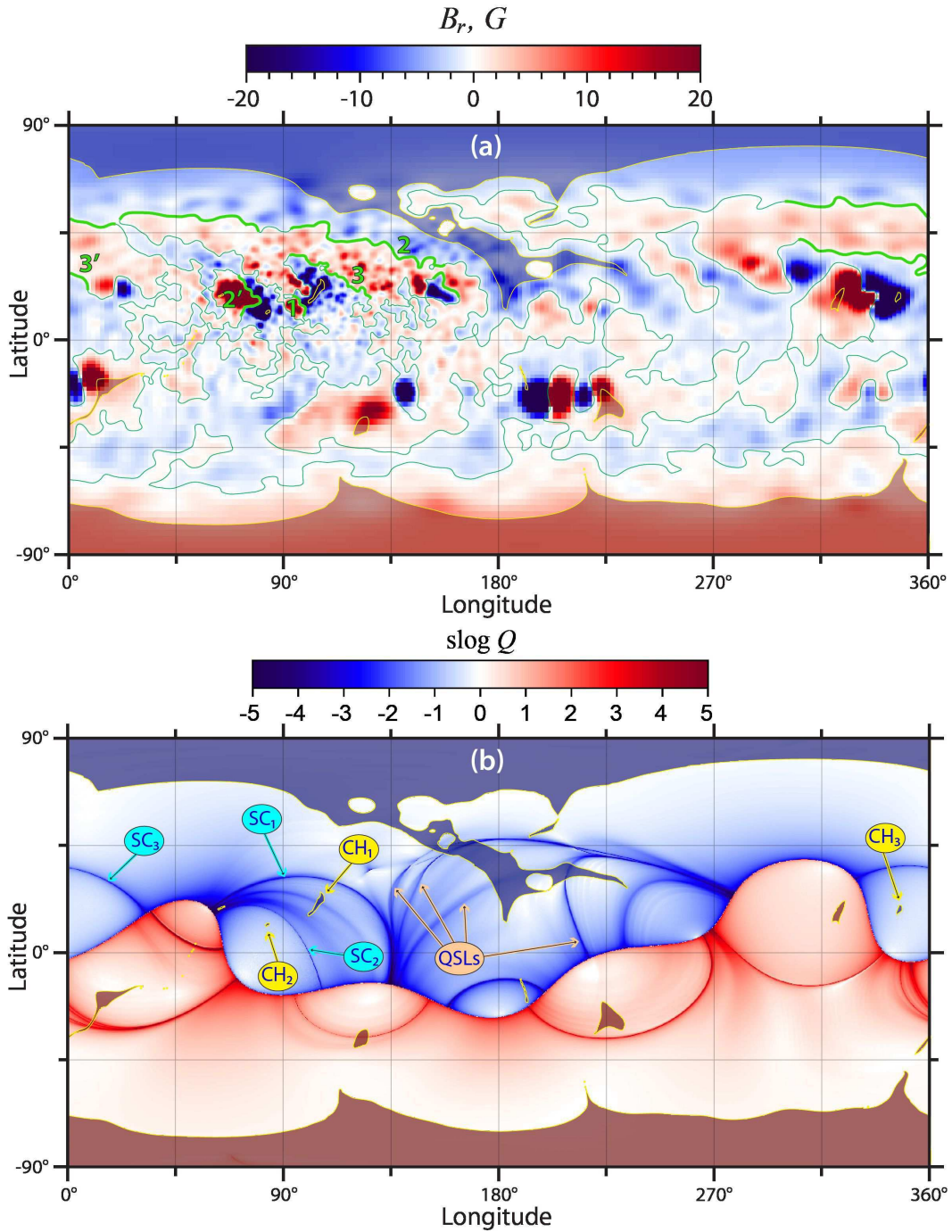}\\
\caption{Map of $B_r$ (a) used as a photospheric boundary condition for our PFSS model of the 2010 Aug 1--2 magnetic field and map of $\slog Q$ for this model at the source surface (b) with superimposed (semi-transparent) photospheric map of coronal holes (shaded either in dark red ($B_r>0$) or dark blue ($B_r<0$) and outlined in yellow). 
Thin (green) lines represent the photospheric polarity inversion line, whose thick segments designate the location of the filaments, part of which are numbered in the order they erupted.
Yellow balloons indicate the coronal holes involved in the eruptions; cyan balloons indicate source-surface footprints of the separatrix curtains of these pseudo-streamers.
	\label{f:maps1} }
\end{figure}

\begin{figure}[]
\epsscale{0.85}
\plotone{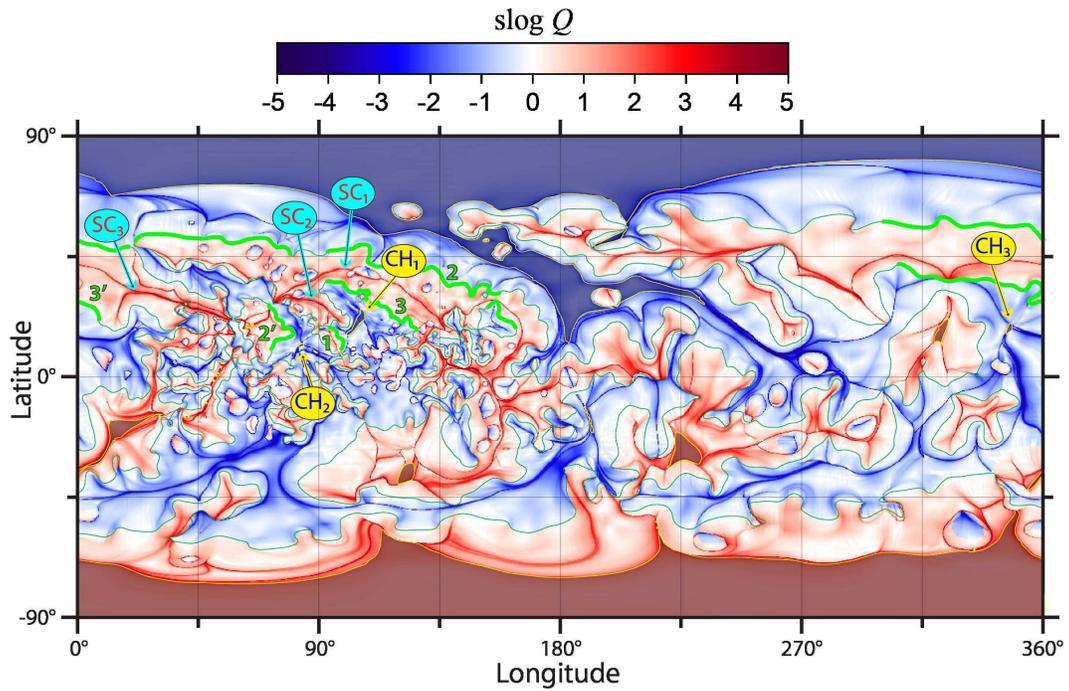}\\
\caption{Map of $\slog Q$ for our PFSS model at the photosphere with superimposed (semi-transparent) photospheric map of coronal holes and the photospheric polarity inversion line, both shown in the same way as in Figure \ref{f:maps1}. 
Yellow balloons indicate the coronal holes of the pseudo-streamers involved in the eruptions; cyan balloons indicate photospheric footprints of the separatrix curtains of these pseudo-streamers.
	\label{f:maps2} }
\end{figure}

The pseudo-streamers we described above (see Figures \ref{f:SC1}--\ref{f:SC3}) illustrate this fact conclusively.
All these cases were characterized by disconnected coronal holes \CH{1}, \CH{2}, and \CH{3} (Figure \ref{f:maps1}), each of which merges with an adjacent coronal hole at the height of the basic null point of the corresponding pseudo-streamer.
At heights where the magnetic field becomes completely open, the corresponding separatrix curtains \SC{1}, \SC{2}, and \SC{3} serve as interfaces between the holes.
Note also that their footprints appear at the source surface as very sharp high-$Q$ lines, whose ends are joined to the null line of the magnetic field (Section \ref{s:hQl}).

Of course, this does not exclude the possibility for different parts of photospheric open-field regions to be connected with each other through narrow corridors.
Several examples of such corridors are also seen in our northern polar coronal hole (Figure \ref{f:maps1}).
They imply the appearance of QSLs in the open field, as proposed first by \citet{Antiochos2007}, and whose transformation into separatrix curtains and back to QSLs have been described at length by \citet{Titov2011}.
As already pointed out in Section \ref{s:hQl}, such QSLs appear at the source-surface as high-$Q$ lines with a smooth distribution of $Q$ across their widths  (Figure \ref{f:maps1} (b)).
Just as in the case of separatrix curtains, these high-$Q$ lines are joined at both ends to the null line of the magnetic field.

Thus, in both the case of truly disconnected and connected coronal holes, interpreted as two-dimensional photospheric regions, their mapping to the source surface connects to the null line of the helmet streamer.
This is in contrast to the interpretation of \citet{Crooker2012}, who regarded this property of the field line mapping as evidence of the connectivity of coronal holes at the photospheric level.
Moreover, we think that the V-shaped coronal hole they interpreted as connected in the photosphere is actually disconnected, as our earlier study of the same case indicates in the framework of the global MHD model \citep{Titov2011}.
This particular example shows that when coronal holes are connected in three dimensions it does not necessarily imply that they are connected in the photosphere too.

It remains to be studied how numerous the above open-field QSLs are, compared to separatrix curtains, in magnetic configurations with a realistically high resolution.  Note that by definition they both belong to the S-web.
In the $\slog Q$-distribution at the source surface, the S-web appears as a network of high-$Q$ arcs connected to the null line of the helmet streamer (Figure \ref{f:maps1} (b)).
The width in latitude of the S-web at this surface is a well-defined quantity, because its value is uniquely related to the open photospheric flux that is (nearly) disconnected from the main bodies of the polar coronal holes.
It is unlikely that this flux, and hence the width of the corresponding S-web, will significantly change if one further increases the resolution of the input magnetic data and the corresponding PFSS model.

This conclusion is in contrast with the statement of \citet{Wang2012} that the S-web will extend to the polar region if one resolves its small parasitic polarities.
Each such polarity will, indeed, bring additional (quasi-)separatrix structures into the open-field regions.   
However, in contrast to the separatrix curtains of pseudo-streamers, these structures will, first, have a much smaller angular size
and, second, will not criss-cross the S-web, but rather stay mostly isolated from it.
Since the quasi-separatrix structures arising from parasitic polarities in polar coronal holes have different geometrical sizes and structural properties, their physical properties are also likely to be different.
Therefore, they have not been included into the definition of the S-web \citep{Antiochos2011}, regardless of the fact that the polar plumes associated with these parasitic polarities might appear similar to pseudo-streamers observationally.

\begin{figure}[h]
\epsscale{1.0}
\plotone{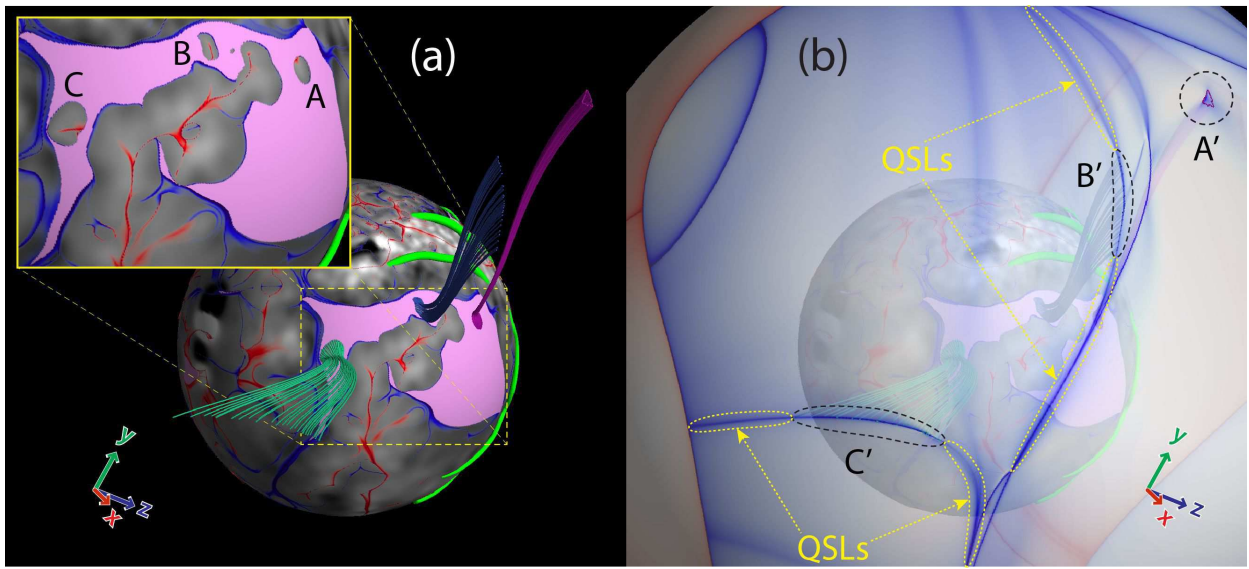}\\
\caption{Structure of magnetic field lines near parasitic polarity regions A, B, and C embedded into the northern coronal hole (a) and location of their footprints A$^\prime$, B$^\prime$, and C$^\prime$ at the source surface (b), where a (semi-transparent) $\slog Q$ distribution is also displayed.
The high-$Q$ lines encircled by dashed (yellow) lines correspond to the footprints of QSLs that originate in the photospheric open-field corridors adjacent to these polarities. 
	\label{f:plums} }
\end{figure}

To clearly make this point, Figure \ref{f:plums} shows what happens around three small parasitic polarities (A, B, and C) embedded into the northern coronal hole.
Panel (a) depicts three sets of open field lines that start very close to the oval high-$Q$ lines bordering the closed magnetic flux of these polarities.
Panel (b) shows their source-surface footpoints  A$^{\prime}$, B$^{\prime}$, and C$^{\prime}$, indicating that such field lines hit the boundary far away from the null line.
Thus, their behavior indeed differs from that of the field lines belonging to the separatrix curtains we described previously.

In particular, as stated above, for polarities that are far from the main border of their surrounding coronal hole, such as A, their signature at the source surface A$^{\prime}$ is completely isolated from the S-web.
Polarities B and C, however, are much closer to the coronal-hole border; they are detached from it by only a relatively narrow open-field corridor.  As expected, the field lines starting in these corridors form QSLs whose footprints at the source surface adjoin on each side of their respective footprints B$^{\prime}$ and C$^{\prime}$ (as shown in Figure \ref{f:maps1} (b)).  The high-$Q$ lines resulting from these merged QSLs would appear, at first sight, to form arcs whose ends join the null line of the helmet streamer.  However, we would argue that these ``arcs'' do not genuinely belong to the S-web because these segments have rather low values of $\log Q$ ($\lesssim 1.5$).
In summary, we have argued that the addition of small parasitic polarities in polar coronal holes would not contribute to the S-web significantly, if at all.  We intend to test this conjecture in future work by explicitly calculating the contribution of parasitic polarities in high-resolution PFSS models.

These considerations help us to predict how our S-web will change with increasing resolution of the input magnetic data and the corresponding PFSS model.
First, increased resolution will cause additional fragmentation of the disconnected coronal holes, while leaving their total magnetic flux approximately unchanged.  
Our analysis suggests that this will increase the number of cells and high-$Q$ lines in the S-web, but will not substantially increase its width in latitude at the source surface.

Depending on the strength of the parasitic polarities introduced when going to higher resolution, and their positions in coronal holes, the separatrix structure enclosing these polarities can be of two types.
First, it can be just a single {\hO bald patch} separatrix surface, as in our examples shown in Figure \ref{f:plums}.
This structure contains no null points in the corona, but nevertheless it completely separates the closed flux of the parasitic polarity from the surrounding magnetic field  \citep{Bungey1996, Meuller2008}.
Second, it can also be a more familiar structure with a dome-like fan surface and spine line across both coming out from a single null point and surrounded by QSLs \citep{Masson2009}.  

These two types of separatrix structures are similar in that their (quasi)-separatrix field lines do not fan out in the open field region as much as they do in pseudo-streamers.
The perturbation of such a structure due to local flux emergence or photospheric motion causes formation of a current sheet and reconnection, both localized in a small region near the corresponding bald patches or null points.
This process can be considered as a mechanism for producing coronal plumes or ``anemone" jets in polar coronal holes \citep{Moreno-Ins2008, Meuller2008, Pariat2010}.

The pseudo-streamers are structurally very different: As shown above, they contain several separators, two of which are open, while the others are closed.
An emergence, submergence, and/or displacement of photospheric flux concentrations in the lobes of pseudo-streamers, and in their surrounding, must lead to the formation of current sheets along the separators closest to the source of the perturbations.
Since current sheets form along the entire length of separators, the related reconnection processes proceed similarly \citep{Parnell2010}.
This indicates that reconnection in pseudo-streamers and coronal plumes might have quite different characteristics, which additionally substantiates the original definition of the S-web. 

The open separators are lines at which the open and closed magnetic fields become in contact with each other.
They appear to be the longest separators in the pseudo-streamers, so most of the interchange reconnection must occur along them.
How does it proceed in the presence of {\hO multiple closed separators}, all connected together with the basic nulls of the pseudo-streamers? 
This question is of particular importance for understanding the physics of pseudo-streamers and has never been investigated before, because their topological structure was unknown.
The answer to this fundamental question is crucial to determine if the S-web model can explain the origin of the slow solar wind.
Therefore, it ought to be the focus of the future studies, with special emphasis on the processes that occur both at open separators and the QSLs associated with open-field corridors. 

The plasma sheets of pseudo-streamers, as observed in the white-light corona, are composed of fine ray-like structures that are presumed to be formed by interchange reconnection at the streamer cusp \citep{Wang2012}. Such an explanation is consistent with our discussion of open separators, except that in our scenario reconnection  occurs along the entire length of these separators rather than just at the mentioned cusp points (which are the footpoints of our open separators at the source surface). In light of our present analysis, the observed ray-like structures are likely a part of the S-web.
For structural features (like separators) to be visible, they have to not only be present, but also perturbed sufficiently  (e.g., by waves or photospheric motions).  Therefore, at any moment in time, only a small fraction of the S-web might be visible in white light.

It should also be emphasized that the S-web model does not assume a priori  that reconnection in pseudo-streamers generates the slow wind in the form of plasma blobs, as it does in helmet streamers \citep{Wang2012}.  In fact, we expect that this process must be so different here that it will directly affect the observational properties of the pseudo-streamers. 
Indeed, in contrast to the helmet streamers, the reconnection in the pseudo-streamers has to occur not in the plasma sheet itself but rather at its edges, where the above open separators are located.  Consequently, the pseudo-streamer material must be replenished, at least in part, by the plasma that flows out from those edges. This process has likely to occur in a sporadic fashion,
namely, each time when the interchange reconnection takes place between open and closed fields. As a result, the respective reconnection outflows have to be modulated accordingly to produce in the pseudo-streamers the mentioned above ray-like rather than blob-like structures.  This consideration shows that, irrespective of its relevance to the problem of the origin of the slow solar wind, the question on how the interchange reconnection modifies the properties of the wind flow in the pseudo-streamers deserves very close attention in the future studies.


\section{SUMMARY}
	\label{s:s}

We have studied the large-scale topology of the coronal magnetic field determined in the framework of a PFSS model  for the time period 2010 August 1--2, when a sequence of sympathetic CMEs occurred.
First, this model was computed from the observed data of the photospheric magnetic field. 
Second, we have calculated high-resolution distributions of the squashing factor $Q$ at the photospheric and source-surface boundaries and at several cut planes across the regions where the CMEs started.
Third, we have developed a special technique for tracing (quasi-)separatrix field lines that pass through the high-$Q$ lines of such distributions.
These tools allowed us to fulfill a comprehensive analysis of the magnetic field structure.

Of particular interest to us were large-scale separatrix surfaces that divide the coronal volume into topologically distinct regions in which the erupting filaments originated.
We have found that four of these five filaments were initially located in the lobes of three pseudo-streamers.
Such lobes are obtained as a result of intersection of curtain-like and dome-like separatrix surfaces of the coronal magnetic field.
The separatrix curtain is a fan separatrix surface associated with a null point that is called basic one and located at a certain height in the corona between two adjacent coronal holes of like polarity.
Such a curtain is formed by open and closed field lines fanning out from the basic null point.
The dome separatrix surfaces are made of two half-domes joined with each other along the spine line of this null point.
The half-domes are formed by the field lines that also fan out either from a bald patch or another null point, which both are located at the flanks of the pseudo-streamer.

In the middle cross-section passing through the basic null of a pseudo-streamer, these separatrix surfaces intersect to produce a \PS-type shape in which the vertical line and arc represent the separatrix curtain and adjoint half-domes, respectively.
Above the half-domes in this cross-section, the curtain separates adjacent coronal holes of like polarity and observationally corresponds to the stalk of pseudo-streamers.
At heights below the basic null of the pseudo-streamer, the coronal holes become disconnected by closed magnetic fields rooted in parasitic polarities and separated by the distance equal to the local width of the separatrix half-domes. 

The separatrix surfaces of the pseudo-streamers in the August 1--2 events are located
relatively far from the pre-eruption positions of the filaments, so that
their contributions to the total field and hence their influence on these surfaces must be small.
Therefore, our source-surface model should be sufficiently accurate  to reproduce the large-scale structure of real pseudo-streamers with filaments inside.

The indicated separatrix curtains intersect half-domes along closed separator field lines, or simply separators, that pass through the null points or bald patches at the flanks of the pseudo-streamers.
In addition, these curtains intersect the helmet-streamer separatrix surface twice along open separator field lines, which connect the basic nulls of the pseudo-streamers to streamer-cusp points.
Invoking our recent MHD model of sympathetic eruptions \citep{Torok2011},
we argue that magnetic reconnection at both these types of separators is likely a key process in sympathetic eruptions,
because it controls how magnetic fluxes are redistributed between the lobes of pseudo-streamers during eruptions.
It has been demonstrated here that the configuration
which harbored the first three erupting filaments had a similar magnetic topology as was assumed in that model.
Thus, the present topological analysis of the PFSS background field substantiates the previous assumptions on the initial configuration in  \citet{Torok2011}.

Here we proceeded with a generalization of this earlier proposed scenario,
by noticing, first, that the indicated separators in our configuration form a huge chain that traverses through all three pseudo-streamers involved in the eruptions.
We have qualitatively explained how a single eruption at one end of such a separator chain can trigger a whole sequence of eruptions.

We have also discussed the implications our obtained results for the S-web model of the slow solar wind by emphasizing those issues that have not been well understood so far.
First, we have demonstrated how the pseudo-streamer structure accommodates disconnection and merging of two coronal holes, respectively, below and above the basic nulls of the pseudo-streamers.
Second, we have explained the differences in magnetic topology between pseudo-streamers and separatrix structures enclosing small parasitic polarities in the polar coronal holes
and discussed why such structures were not included in the original definition of the S-web.
Third, we have emphasized that the sources of the slow solar wind most likely reside both at the separators of pseudo-streamers and QSLs originated in narrow photospheric corridors of the open magnetic field.



\acknowledgments

{\hG

The contribution of V.S.T., Z.M., T.T., and J.A.L. was supported by NASA's HTP, LWS, and SR\&T programs,
SHINE under NSF Grant AGS-1156119, CISM (an NSF Science and Technology Center), and a contract from Lockheed-Martin to Predictive Science, Inc.; O.P. was supported by NSF Grant ATM-0837519 and NASA Grant NNX09AG27G.

}



%

%




\clearpage

\end{document}